
\magnification=1200
\hsize 15true cm \hoffset=0.5true cm
\vsize 23true cm
\baselineskip=15pt

\font\grande=cmr10 scaled \magstep4
\font\medio=cmr10 scaled \magstep2
\outer\def\beginsection#1\par{\medbreak\bigskip
      \message{#1}\leftline{\bf#1}\nobreak\medskip\vskip-\parskip
      \noindent}

\def\laq{\raise 0.4ex\hbox{$<$}\kern -0.8em\lower 0.62
ex\hbox{$\sim$}}
\def\gaq{\raise 0.4ex\hbox{$>$}\kern -0.7em\lower
0.62 ex\hbox{$\sim$}}

\def \pa {\partial}
\def \ra {\rightarrow}

\def \fb {\overline \phi}
\def \rb {\overline \rho}
\def \pb {\overline p}
\def \pr {\prime}

\def \ti {\tilde}
\def \t {\theta}
\def \la {\lambda}

\def \b {\beta}
\def \a {\alpha}
\def \ap {\alpha^\prime}

\def \Ga {\Gamma}
\def \ga {\gamma}
\def \sg {\sigma}
\def \da {\delta}

\def \r {\rho}

\def \om {\omega}

\def \noi {\noindent}

\def\sqr#1#2{{\vcenter{\hrule height.#2pt\hbox{\vrule width.#2pt
height#1pt \kern#1pt\vrule width.#2pt}\hrule height.#2pt}}}

\def\lsim{\mathrel{\rlap{\lower4pt\hbox{\hskip1pt$\sim$}}
    \raise1pt\hbox{$<$}}}         
\def\gsim{\mathrel{\rlap{\lower4pt\hbox{\hskip1pt$\sim$}}
    \raise1pt\hbox{$>$}}}         

\nopagenumbers
\line{\hfil  CERN-TH/95-40}
\line{\hfil  DFTT-16/95}
\vskip 1.5 cm
\centerline {\grande  Evolution of a String Network}
\vskip 0.5 true cm
\centerline{\grande in Backgrounds with Rolling Horizons}

\vskip 1true cm
\centerline{M. Gasperini, M. Giovannini}
\centerline{\it Dipartimento di Fisica Teorica, Universit\`a di
Torino,}
\centerline{\it Via P.Giuria 1, 10125 Turin, Italy,}
\centerline{K. A. Meissner}
\centerline{\it Institute of Theoretical Physics, ul. Hoza 69, 00-681
Warsaw, Poland}
\centerline{and}
\centerline{G. Veneziano}
\centerline{\it TH. Division, CERN, CH-1211 Geneva 23, Switzerland}

\vskip 1true cm
\centerline{\medio Abstract}

\noindent
We discuss the temporal variation of the equation of state of
 a classical string network, evolving in a background in which
the Hubble radius $H^{-1}$ shrinks to a minimum and then re-expands
to infinity. We also present a method to look for self-consistent
non-vacuum string backgrounds, corresponding to the simultaneous
solution
of the gravi-dilaton background equations and of the string
equations of motion.

\vskip 1.5true cm
\centerline{---------------------------}
\centerline {To appear in the {\bf "String gravity and physics at the
Planck energy scale "} }
\centerline { (World Scientific, Singapore, 1995)}

\vskip 1.5 cm
\noindent
CERN-TH/95-40

\noindent
February 1995
\vfill\eject

\footline={\hss\rm\folio\hss}
\pageno=1
\centerline{\bf EVOLUTION OF A STRING NETWORK}
\centerline{\bf IN BACKGROUNDS WITH ROLLING HORIZONS}
\bigskip
\centerline{M. Gasperini, M. Giovannini}
\centerline{\it Dipartimento di Fisica Teorica, Universit\`a~di Torino,}
\centerline{\it Via P.Giuria 1, 10125 Turin, Italy,}
\centerline{K. A. Meissner}
\centerline{\it Institute of Theoretical Physics, ul. Hoza 69, 00-681
Warsaw, Poland}
\centerline{and}
\centerline{G. Veneziano}
\centerline{\it TH. Division, CERN, CH-1211 Geneva 23, Switzerland}

\bigskip
\centerline{ABSTRACT}
\midinsert
\narrower
\noi
We discuss the temporal variation  of the equation of state of
 a classical string network, evolving in a background in which
the Hubble radius $H^{-1}$ shrinks to a minimum and then re-expands
to infinity. We also present a method to look for self-consistent
non-vacuum string backgrounds, corresponding to the simultaneous
solution
of the gravi-dilaton background equations and of the string
equations of motion.
\endinsert
\vskip 1 cm
\noi
{\bf 1. Introduction.}

\noi
Extensive studies of string propagation in cosmological
backgrounds [1-4], have shown that
the qualitative behaviour of
classical strings  depends dramatically on the
 presence of event horizons in the given background
geometry. In this case  it is possible  to define
two physically distinct regimes for the classical solutions,
 according to whether the proper size $L$ of a
 string, $L \sim \alpha' E$, is smaller or
larger than the size of the horizon $L_h$. It turns out
that, for $L<<L_h$, the solutions are oscillatory
with constant proper size,
  while the regime
$L>>L_h$ is characterized by non-oscillatory  solutions, in which the
string proper size evolves in time like
the scale factor, $a(t)$, of the background geometry (see also [5]).

This latter regime, first discussed in the context of a
de Sitter-type manifold [1,2,5], is of crucial importance for all
those
metric backgrounds in which the size of the event horizon
 shrinks to zero. Classically, in such geometries,
 strings of any size  asymptotically become larger than the
horizon. The classical description, however, is expected to break
down at some point. Thus,
 while the string proper size is  expected to be always larger than
$\lambda_s = \sqrt{\alpha' \hbar}$ as a result of the
uncertainty principle  [6], classical geometries
 with horizon size
 smaller than $\lambda_s$ are also expected to suffer
 large quantum corrections.
Since we want to keep our discussion here  entirely classical
 we shall use $\lambda_s$ as a short distance cut-off  both for the
string
size distribution and for the horizon of the background geometry.

The shrinking of the event horizon is
typical of inflationary scenarios (also called "pre-big-bang"
scenarios) which follow
 from the low energy string effective action [7-8],
 the accelerated evolution of the metric being driven by the
kinetic energy of the dilaton field. A question which arises
naturally
in that context is whether the back reaction of string matter,
evolving in
the presence of shrinking horizons, could lead to
 a string-plus-dilaton-driven inflationary
scenario which solves simultaneously, and
self-consistently, both the background field equations and the string
equations of motion (very recent work on similar issues can be
found in [9]).

In order to avoid confusion, we  recall that the
shrinking of the event horizon can occur in two types of
backgrounds [8], characterized, respectively, by a superinflationary
accelerated expansion,
$$
\dot a >0,\,\,\,\,\,\,\, \ddot a >0, \,\,\,\,\,\, \, \dot H >0
\eqno(1.1)
$$
or by an accelerated contraction
$$
\dot a <0,\,\,\,\,\,\,\, \ddot a <0, \,\,\,\,\,\, \, \dot H <0
\eqno(1.2)
$$
of the scale factor $a(t)$. Here $H=\dot a/a$, and a dot denotes
differentiation with respect to  cosmic time $t$.

Consider, in fact, a spatially flat homogeneous and isotropic
 metric,
 $$ g_{\mu\nu}=diag (1, -a^2 \da_{ij}),
 \,\,\,\,\,\,\,\,\, i,j=1,...,d \geq 3 \eqno(1.3)
$$
parameterized by
$$
a(t)\sim (-t)^\a \eqno(1.4)
$$
for $t\ra 0_-$. The proper size of the event horizon
$$
L_h(t)=a(t)\int_t^0 dt'a^{-1}(t') \eqno(1.5)
$$
is finite, and shrinks linearly in cosmic time, $L_h(t)\simeq (-t)$,
 for
all $\a<1$. The negative part of this range ($\a <0$) corresponds to
superinflation, eq.(1.1), while the positive part ($0<\a <1$)
corresponds to accelerated contraction, eq.(1.2).

Both superinflation and accelerated
contraction provide a representation of the pre-big-bang scenario
[7,8], the former  in the so-called string  frame,
(in which weakly coupled test strings move along geodesic surfaces
of the background metric), the latter in the Einstein frame (in which
the dilaton is minimally coupled to the metric and the corresponding
effective action is diagonalized in the standard canonical form).

Obviously, the abovementioned phenomena, being of purely geometrical
nature, refer just to the
 string-frame metric (the dilaton controls instead
the strength of the mutual
 interaction among strings,
 assumed to be negligible in the following).
For all these reasons we will limit
our attention to a background which describes
(in the string frame) an initial
superinflationary era followed by
the time-reversed version of an
accelerated contraction: the latter phase describes nothing more
than the
standard (non-inflationary) decelerated expansion ($\dot a >0$,
$\ddot a <0$) of the present epoch. It was shown in [10] that a
transition between the two epochs cannot occur in the low curvature
regime: an intermediate genuinly stringy epoch is needed as well.

Consider now the string equations of motion which, in the gauge in
which the world-sheet metric is conformally flat, can be written
explicitly as [1-3]
$$
{d^2x^\mu \over d\tau^2} - {d^2x^\mu \over d\sg^2} +
\Ga_{\a\b}^\mu
({dx^\a \over d\tau} + {dx^\a \over d\sg} )({dx^\b \over d\tau} -
{dx^\b \over d\sg} )=0
$$
$$
g_{\mu\nu}({dx^\mu \over d\tau} {dx^\nu \over d\tau} +
{dx^\mu \over d\sg} {dx^\nu \over d\sg} )=0,\,\,\,\,\,
g_{\mu\nu}{dx^\mu \over d\tau} {dx^\nu \over d\sg} =0 \eqno(1.6)
$$
where $x^\mu (\sg, \tau)$ are the string coordinates, $\tau$ and
$\sg$ the usual world-sheet time and space variables, and
$\Ga_{\a\b}^\mu$ is the Christoffel connection for the background
metric (1.3). When the string proper size is negligible with respect
to
the horizon, the exact solution of these equations can be expressed
as an expansion around the point-particle motion of the string
center of mass [1]. If the horizon is shrinking, however, such an
expansion breaks down, and the explicit solution of eqs.(1.6) shows
that the string comoving size becomes frozen, asymptotically, while
the proper size evolves in time like the scale factor, $L(t)\sim
a(t)$
for $L(t)>>L_h(t)\sim (-t)$.

This occurs both for superinflation and for accelerated contraction
[3]. The crucial difference between the two types of background,
however, is that, in an expanding geometry, the asymptotic solutions
of eqs.(1.6) are characterized by
$$
\left|{\pa x^0 \over \pa \tau }\right| >>
\left|{\pa x^0 \over \pa \sg
}\right|,\,\,\,\,\,\,\,\,\,\,\,\,\,\,\,\,\,\,
\left|{\pa x^i \over \pa \tau }\right|<<
\left|{\pa x^i \over \pa \sg }\right| \eqno(1.7)
$$
while, in the contracting case, they satisfy
$$
\left|{\pa x^0 \over \pa \tau }\right| >>
\left|{\pa x^0 \over \pa \sg
}\right|,\,\,\,\,\,\,\,\,\,\,\,\,\,\,\,\,\,\,
\left|{\pa x^i \over \pa \tau }\right|>>
\left|{\pa x^i \over \pa \sg }\right| \eqno(1.8)
$$
As a consequence, the string energy-momentum tensor
$$
T^{\mu\nu}(x)= {1\over \pi \ap \sqrt{|g|}}\int d\sg d\tau (
{dx^\mu \over d\tau} {dx^\nu \over d\tau}-
{dx^\mu \over d\sg} {dx^\nu \over d\sg}) \da^{d+1}
 (x-x(\sg, \tau) )\eqno(1.9)
$$
($(\a')^{-1}$ is the string tension) satisfies respectively the
asymptotic conditions
$$
T_0\,^0\simeq\sum_iT_i\,^i ,  \,\,\,\, \a<0,
\,\,\,\,\,\,\,\,\,\,\,\,\,\,
T_0\,^0\simeq-\sum_iT_i\,^i ,  \,\,\,\,  0<\a<1,
\eqno(1.10)
$$
In the perfect fluid approximation, $T_0\,^0=\r $, $T_i\,^j
=-p\da_i^j$, one thus obtain for a diluted gas of classical strings
the
effective equations of state [3]
$$
{p\over \r}\simeq -{1\over d},  \,\,\,\, \a<0,
\,\,\,\,\,\,\,\,\,\,\,\,\,\,
{p\over \r}\simeq {1\over d},  \,\,\,\,  0<\a<1,
\eqno(1.11)
$$
The relationship between the two regimes can be easily understood in
terms
of the so-called scale-factor duality property of string motions
in cosmological backgrounds [7].

We note, for comparison, that in the opposite regime $L(t)<<L_h(t)$
the string solutions are instead characterized by the condition
$$
\left|{\pa x^0 \over \pa \tau }\right| >>
\left|{\pa x^0 \over \pa \sg
}\right|,\,\,\,\,\,\,\,\,\,\,\,\,\,\,\,\,\,\,
\left|{\pa x^i \over \pa \tau }\right| \simeq
\left|{\pa x^i \over \pa \sg }\right| \eqno(1.12)
$$
which implies
$$
T_0\,^0 >>\sum_iT_i\,^i ,  \,\,\,\,  \,\,\,\,\,\,\,\,\,\,\,
{p\over \r}\simeq  0  \eqno(1.13)
$$

As it is clear from the way they are obtained,  the effective
equations of state (1.11) and (1.13), characterized by a constant
ratio
$\ga=p/\r$, are only valid asymptotically when strings are,
respectively, far outside or well inside the horizon. The purpose of
this paper is to discuss instead the time-variation of the ratio
$p/\r$
(for a gas of classical strings) versus the time-variation of the
event
horizon, considering in particular a background in which the Hubble
radius $H^{-1}$ shrinks to a minimum and then re-expands to
infinity. To this aim we shall propose in Section 2 an effective
evolution equation for the density of strings of arbitrary proper
size, and we shall present in Section 3 some examples  of explicit
solution of such equation. A possible approach to the problem of
obtaining simultaneous and self-consistent solutions to the system of
background field equations and string equations of motion will be
discussed in Section 4. Our main results will be finally summarized
in
Section 5.
\vskip 2 cm
{\bf 2. Evolution equation for the string distribution.}

Let us consider a cosmological metric background of the type (1.3)
which, starting from an asymptotically flat initial state at
$t=-\infty$, evolves towards a high curvature regime [7-8]. The
curvature reaches at $t=0$ a maximum scale of order
$\la_s^{-1} \laq {\ell}_p^{-1}$ (where ${\ell}_p$ is the
Planck length), and then decreases approaching zero as $t\ra
+\infty$. In such a background, the Hubble length  $D(t)\equiv
H^{-1}$
(which we shall assume here to be always positive) shrinks to a
minimum $D(0)\simeq \la_s$, and then re-expands to infinity.

Consider, in this background, a string of initial proper size
$L_{-\infty}$. According to the results reviewed in the previous
Section [1-5], the string size stays constant inside the horizon, and
evolves like $a(t)$ outside the horizon. The time-dependence of the
string proper size $L(t)$ can thus be fixed by the equation
$$
L(t)=L_{-\infty}\theta(-t)\theta(D(t)-L(t))+L_{-\infty}\theta(L(t)-D(t))
{a(t)\over a(t_{out})}+
$$
$$
+L_{+\infty}\theta(D(t)-L(t))\theta(t) \eqno(2.1)
$$
where $\t$ is the Heaviside step function,
$$
L_{+\infty}=L_{-\infty}{a(t_{in})\over a(t_{out})} \eqno(2.2)
$$
and $t_{out}$ and $t_{in}$ are, respectively , the times
at which the string leaves and re-enters the horizon, determined by
the conditions $D(t_{out})=L_{-\infty}$, $D(t_{in})=L_{+\infty}$.

Consider next, in the same background, a network of
$N$  such non-interacting strings, occupying initially a certain
comoving volume. Let us denote by
 $L^{i}(t)$ ($i=1,2,...,N$) the individual proper sizes at time $t$.
The number
distribution $n(L,t)$ of strings of proper size $L$, at time $t$,
such
that $N=\int n dL$, is obviously given by
$$
n(L,t)=\sum_{i=1}^{N}\delta(L-L^{i}(t))\eqno(2.3)
$$
We now differentiate this distribution with respect to $L$ and $t$
and
take into account that, from eq.(2.1),
$$
\dot L^{(i)}=H(t)\theta(L^{(i)}(t)-D(t)) L^{(i)}\eqno(2.4)
$$
We thus find that the number distribution of a diluted gas of
classical strings, in the considered background, must satisfy the
approximate evolution equation
$$
{\partial n(L,t)\over \partial t}+ H(t){\partial\over \partial
L}[n(L,t)L\theta(L-D)]=0\eqno(2.5)
$$
Equivalently, by using $a$ instead of $t$ as evolution parameter,
$$
a{\partial n(L,a)\over \partial a}+ {\partial\over \partial
L}[n(L,a)L\theta(L-D)]=0\eqno(2.6)
$$

The general solution of this equation can be conveniently
expressed in terms of the initial string distribution,
$n_{-\infty}(L)$,
as
$$
n(L,a)=n_{-\infty}(L)\theta(D-L) +{1\over a} h({L\over a})\theta(L-D)
\eqno(2.7)
$$
where
$$
h(\xi)=a(\xi)n_{-\infty}(D(\xi))\left[1+{\partial \ln
a(\xi)\over \partial \ln   H(\xi)}\right]^{-1}     \eqno(2.8)
$$
($\xi= {D\over a}$). Once the background $a(t)$ and the initial string
distribution $n_{-\infty}$ are given one can thus compute all
physical observables, and in particular the energy density of strings
which are smaller and larger than the horizon (which we shall call
for simplicity stable and unstable, respectively).

By recalling that the string energy is proportional to the proper
size
$L$, the energy  of stable and unstable strings in our ensemble can
be
estimated, respectively, as
$$
E_s=\int d^3x \sqrt{|g|}\r_s= {1\over \pi \a'}\int L n(L,a)\t (D-L)
dL
$$
$$
E_u=\int d^3x \sqrt{|g|}\r_u= {1\over \pi \a'}\int L n(L,a)\t (L-D)
dL
\eqno(2.9)
$$
(the integral extends of course over the whole range of $L$, from
the minimal allowed proper size $\la_s$ up to infinity). The
corresponding pressure, on the other hand, is given asymptotically
by the effective equations of state (1.13) and (1.11), respectively.
The
time evolution of the ration $\ga =p/\r$ for the full string
distribution, therefore, can be approximated as
$$
\ga(t)=\pm{E_u\over d E}=\pm {1\over d}{\int_{\la_s}^{\infty} L n\t
(L-D) dL \over \int_{\la_s}^{\infty} L n d L } \eqno(2.10)
$$
(here $E=E_s+E_u$, and the sign of $\ga$ depends on the type of
background, according to eq.(1.11)).

The time-dependence of $\ga$ determined in this way relies on the
evolution equation (2.6), obtained in the sudden approximation in
which the transition from the stable to unstable regime is very
roughly parameterized by the step function $\t (L-D)$. Such a
transition, however, could be be approximated in a continuous way
by introducing an appropriate function $\t (L,a)$ which goes to $0$
for $LH<<1$ and to $1$ for $LH>>1$, and which interpolates smoothly
between strings smaller and larger than the horizon. The evolution
equation for $n$, written in terms of such a smooth function,
$$
a{\partial n(L,a)\over \partial a}+ {\partial\over \partial
L}[n(L,a)L\theta(L,a)]=0\eqno(2.11)
$$
has then the general solution
$$
n(L,a)=-\sg (z) {\pa z \over \pa L}\eqno(2.12)
$$
where $\sg$ is an arbitrary function of $z$, and $z(L,a)$ is a
particular solution of the differential equation
$$
a{\pa z \over\pa a}  +\theta(L,a)L{\pa z\over \pa L}=0\eqno(2.13)
$$

In the following Section we shall provide examples of
solution of the evolution equation for the string number
distribution,
considering both the sudden approximation, eq.(2.6), and the
continuous one, eq.(2.11).
\vskip 2 cm

{\bf 3. Examples of exact solutions.}

We shall consider first  an accelerated background
corresponding to a phase of pre-big-bang evolution, [7-8],
parameterized by the scale factor (1.4) with $\a <1$, and $-\infty
\leq
t \leq -t_1$, where $t_1\simeq \la_s$. We shall assume, moreover, a
power-law behavior for the initial string distribution,
$$
n_{-\infty}(L) = \Lambda ^{\beta -1} L^{-\beta}\eqno(3.1)
$$
Here $\b>2$ for the convergence of the energy integral $\int ^\infty
L n d L$, and $\Lambda$ is an appropriate length parameter, related to
the
total string number $N$ by
$$
N= \int_{\la_s}^{\infty} n_{-\infty}(L) d L = {1\over \b -1}\left (
\Lambda\over \la_s \right)^{\b -1} \eqno(3.2)
$$

In this background $(\pa \ln a / \pa \ln H) = -\a$, and from
the general solution (2.7) we find that the
energy of stable and unstable strings is given, respectively, by
$$
\pi \a' E_s=\int_{\la_s}^{D}L n dL ={\Lambda^{\b-1}\over \b-2}\left(
\la_s^{2-\b} - D^{2-\b}\right)
$$
$$
\pi \a' E_u=\int_{D}^{\infty}L n dL ={\Lambda^{\b-1}\over \b +\a-2}
 D^{2-\b}
\eqno(3.3)
$$
(we have supposed $\b +\a >2$). Note that $\dot E =HE_u$, in
agreement with the covariant conservation of the stress tensor
($\dot \r +dH(\r+p)=0$), and with the fact that only unstable strings
contribute to the total pressure, with $p=-\r/d$.
As a consequence of eqs.(3.2), (3.3), the ratio $\ga(t)$ evolves in
time
(according to eq.(2.10)) as
$$
\ga(t) = \pm {1\over d}{\b-2 \over (\b+\a -2)|H\la_s|^{2-\b} -\a},
\,\,\,\,\, \b>2,\,\,\,\,\, \b+\a >2 \eqno(3.4)
$$
Asymptotically,
$$
\ga =\pm {1\over d} {\b-2\over \b+\a-2}
      |H\la_s|^{\b-2}, \,\,\,\,\,\,\,\,\,\, \, |H\la_s| \ra 0
\eqno(3.4a)
$$
$$
\ga =\pm {1\over d} ,
     \,\,\,\,\,\,\,\,\,\,\,\,\,\,\,\,\, \, |H\la_s| \ra 1
\eqno(3.4b)
$$

Here $sign\{\ga\}=sign \{\a\}$, so that in case of superinflationary
expansion $\ga$ approaches zero from negative values as $t\ra
-\infty$, while in case of accelerated contraction $\ga$ approaches
zero from positive values. If we consider a background which
evolves from a phase of pre-big-bang accelerated expansion,
$$
a(t) \sim (-t)^\a, \,\,\,\,\,\,\,\,\,\, \a <0,\,\,\,\,\,\,t<0
\eqno(3.5)
$$
to a phase of standard decelerated expansion,
$$
\ti a(t) \sim t^\da, \,\,\,\,\,\,\,\,\,\, 0<\da <1,\,\,\,\,\,\,t>0
\eqno(3.6)
$$
we must expect, therefore, a change of trend in the equation of
state,
corresponding to the evolution of $\ga$ which starts from $0$ at
$t=-\infty$, reaches a negative minimum of order unity at $t=0$
(when
$H_1\simeq \la_s^{-1}$), and then starts to increase towards
positive values, approaching  zero as $t\ra +\infty$.

In order to describe the time evolution of $\ga$ in the decelerated
regime following the phase of pre-big-bang inflationary expansion,
we write first the general solution for $n$ in the background (3.6)
 and then demand that it joins smoothly to the final
distribution originating
from the
preceding inflationary background (3.5). One easily gets, at $t>0$
$$
n(L, \ti a)=n_{+\infty}(L)\theta(D-L) +{1\over \ti a} h({L\over \ti
a})\theta(L-D) \eqno(3.7)
$$
where $n_{-\infty}$ has been replaced by the final asymptotic
string distribution for $t\ra +\infty$,
$$
n_{+\infty}(\xi) = (1-\da) {h(\xi)\over \ti a (\xi)}\eqno(3.8)
$$
(according to eq.(2.8)), and
$$
h(\xi)=A \xi ^{(\a-\b)/(1-\a)}\eqno(3.9)
$$
Here the $\xi$-dependence of $h$ is obtained from eq.(2.8) written
in the background (3.5), and $A$ is a constant coefficient which we
shall fix by normalizing the total number of strings to that of the
initial distribution, $N= \int n_{-\infty} d L $.

For the decelerated background (3.6) $\xi \sim D^{1-\da}$. From
eq.(3.8) we get, therefore,
$$
n_{+\infty}(L)=c_{+\infty} L^{[\da (\b-1)+\a -\b]/(1-\a)}\eqno(3.10)
$$
where the number $c_{+\infty}$ depends on $A, \a, \b$ and $\da$. On
the other hand, if we start at $t=-\infty$ with the initial
distribution
(3.1), we have
$$
n_{+\infty}(L)dL={c_{+\infty} \over \Lambda^{\b-1}}\left(1-\a\over
1-\da \right) n_{-\infty}(L^{{1-\da \over 1-\alpha}})
 d(L^{{1-\da \over 1-\alpha}})
\eqno(3.11)
$$
so that we can always choose $A$ in such a way that the total
number of strings is conserved in the transition from $a$ to $\ti a
$.
This in agreement with a classical description of string propagation
in curved backgrounds, where any quantum string creation or string
decay process is completely neglected (see [9] for a discussion
which includes the effects of quantum decay through string
splitting).

We note, finally, that for the general solution (3.7) the energy of
stable and unstable strings is given, respectively, by
$$
\pi \a' E_s={c_{+\infty}\over k-2}\left(
\la_s^{2-k} - D^{2-k}\right)
$$
$$
\pi \a' E_u={c_{+\infty}\over k +\da-2}
 D^{2-k}, \,\,\,\,\,\,\,\,\,\,\,\,\,\,\,
k=-{\da (\b-1)+\a -\b \over 1-\a}
\eqno(3.12)
$$
($k>2$ for a finite total energy). The corresponding ratio $\ga (t)$
increases towards zero, for $t\ra +\infty$, as
$$
\ga(t) = -{1\over d}{k-2 \over k +\da  -2}
(H\la_s)^{k-2} \eqno(3.13)
$$
The transition from a phase of shrinking horizons to a phase of
expanding horizons is associated, therefore, to a switching of
$\dot \ga$ from negative to positive values, and to a subsequent
decreasing of $|\ga|$ as more and more strings re-enter the horizon.

This behavior of $\ga$, obtained in the context of the sudden
approximation, can be qualitatively confirmed by using a different
approach based on the smoothed evolution equation (2.11). Consider
in fact the continuous scale factor
$$
a =\left(t+\sqrt{t^2+t^2_1}\over t_1\right)^{1/2}, \,\,\,\,\,\,
D\equiv H^{-1} =  t_1(a^2+a^{-2}), \,\,\,\,\,
-\infty \leq t \leq +\infty \eqno(3.14)
$$
which is  self-dual [7,8], i.e. satisfies $a(t)=a^{-1}(-t)$.
 It connects smoothly
the standard radiation-dominated expansion, $a\sim t^{1/2}$ for
$t\ra +\infty$, to the duality-related inflationary regime, $a\sim
(-t)^{-1/2}$ for $t\ra -\infty$. By choosing
$$
\theta(L,a)={L\over L+H^{-1}}\eqno(3.15)
$$
as interpolating function between the stable ($LH>>1$) and unstable
($LH<<1$) regime, one can easily check that
$$
z(L,a)={2t_1\over L}+\arctan\left(a^2+{L\over 2t_1}\right)-
{\pi\over 2}\eqno(3.16)
$$
is a particular solution of eq.(2.13). We choose, moreover,
$$
\sg (z)= {N \pi z\over \left(z+{\pi \over 2}\right)^3}\eqno(3.17)
$$
as simple example of distribution which leads to a finite total
number of strings and to a finite total energy for $L\ra 0$ and
$L\ra \infty$.

 From the solution (2.12) of the smoothed evolution equation we
obtain, in particular,
$$
N =\int_0^{\infty}dL\ n(L,a) = \int_0^{\infty} \sg (z) d z
$$
$$
\pi \a' E = \int_0^{\infty}dL\ Ln(L,a)
\ra c_1 + c_2 a^2, \,\,\,\,\,\,\,\,\,\,\,\, a\ra 0
$$
$$
\pi \a' E = \int_0^{\infty}dL\ Ln(L,a)
\ra {4 N t_1\over \pi}-{16 N t_1\over \pi^2 a^2}\ln \left(\pi
a^2\over
2\right), \,\,\,\,\,\,\,\,\,\,\,\, a\ra \infty
$$
where $c_1$ and $c_2$ are positive constants. On the other hand, in
$d$ isotropic dimensions, the pressure to energy density ratio for a
perfect gas contained inside a proper volume $V$ can be expressed as
$$
\ga = {p\over \r}= -{V\over E}{\pa E \over \pa V}= -{a\over dE}
{\pa E\over \pa a}\eqno (3.19)
$$
The behavior of $\ga$ over the whole range of $t$ can then be
found by integrating numerically the given string energy
distribution.
For what concerns the asymptotic behavior of $\ga$, however, we
can easily obtain from eqs.(3.18) that
$$
t \ra -\infty, \,\,\,\,\, a\ra 0\,\,\,\,\, \, \Longrightarrow
\,\,\,\,\,\,
\ga \sim - a^2 \sim -{1\over (-t)} \sim -H_\infty
$$
$$
t \ra +\infty, \,\,\,\,\, a\ra \infty\,\,\,\,\, \, \Longrightarrow
\,\,\,\,\,\, \ga \sim -{1\over a^2 }\ln a^2  \sim
-H_\infty \ln H_\infty\eqno(3.20)
$$
where $H_\infty \equiv H(t=\pm \infty)=1/(2|t|) $ (see eq.(3.14).
Therefore, $\ga$ starts decreasing from zero at $t \ra -\infty$,
and then eventually increases to approach zero from negative values
as $t \ra +\infty$,
in full qualitative agreement with our previous conclusions. It
should
be noted, moreover, that $\ga$ is always negative, in agreement
with eq.(3.19) and with the fact that the total string energy is
growing,
$$
\dot E =-H\int_0^\infty L dL {\pa \over \pa L} (nL\theta) =
H \int_0^\infty  dL n L \theta \equiv HE_u >0 \eqno(3.21)
$$
over the whole time range $-\infty \leq t \leq +\infty$.

The initial distributions considered in this section consist,
essentially, of short tests strings.
The mean length of the strings is $O(\lambda_{s})$ in the
case of the distribution (3.2), and $O({t_{1}})$ in the case of the
distribution (3.17). The same results of the previous examples can be
obtained, however, also starting with a distribution of strings
with arbitrary mean length $L_{0}$. In this case, if $L_{0}$ is
sufficiently large compared  to $\lambda_{s}$, then
 the ratio ${p\over \rho}$ approaches the minimum value
$-1\over d$  as soon as D becomes much smaller than $L_{0}$.

Consider for instance an accelerated
expanding background described  by the scale factor (3.5),
and assume the following initial distribution :
$$
n_{-\infty}(L)={ N\over L_{0}} e^{-{L\over L_{0}}} \eqno(3.22)
$$
In this case the mean string length is:
$$
< L > ={{\int_{\lambda_{s}}^{\infty}L n_{-\infty}(L) dL}\over
{\int_{\lambda_{s}}^{\infty} n_{-\infty}(L) dL}}=
 L_{0} + \lambda_{s}   \eqno(3.23)
$$
The energy of stable and unstable strings is then, from the general solution
(2.7),
$$
{1\over N}\pi\alpha^{\prime} E_{s}=\int_{\lambda_{s}}^{D} dL L n(L,a) =
(\lambda_{s} +L_{0}) e^{-{\lambda_{s}\over L_{0}}}
- (D+L_{0})e^{-D\over L_{0}}  \eqno(3.24)
$$
$$
{1\over N}\pi\alpha^{\prime} E_{u} = \int_{D}^{+\infty} dL L n(L,a)=
({D\over L_{0}})^{\alpha} L_{0}
\Gamma\left( 2-\alpha,{1\over L_{0}} (-\alpha)^{-\alpha\over {\alpha -1}}
({D\over a})^{1\over {1-\alpha}}\right) \eqno(3.25)
$$
( $\Gamma$ is the incomplete Euler Gamma function).

In order to give a particularly simple example we will
choose  now an inflationary background with $\alpha =-1$,
so that $E_{u}$ reduces to:
$$
{1\over N}\pi\alpha^{\prime} E_{u}=
L_{0} ({L_{0}\over D})[({D\over L_{0}})^2 +
2 ({D\over L_{0}}) +2] e^{-D\over L_{0}}   \eqno(3.26)
$$
and the ratio $\gamma(t)$ becomes, according to equation (2.10),
$$
\gamma(t) = -{1\over d}  {1\over (1+  {{({\lambda_{s}\over L_{0}} +1)
e^{-{\lambda_{s}\over L_{0}}}
-({D\over L_{0}} +1 ) e^{-{D\over L_{0}}}}\over {({D\over L_{0}} +2
{L_{0}\over D} +2) e^{-D\over L_{0}} }})}        \eqno(3.27)
$$
{}From this equation we can easily check that if $\lambda_{s}\simeq L_{0}$
then $\gamma\rightarrow 0$ for  $D>>\lambda_{s}$
and $\gamma \rightarrow -{1\over d}$ for $D\rightarrow \lambda_{s}$,
in agreement with our previous examples.
If, on the contrary, we assume an initial distribution with
$L_{0}>>\lambda_{s}$ then we have $\gamma \simeq -{1\over d}$
as soon as $D<<L_{0}$.
This means that, for a gas of sufficiently long strings,
the minimal negative pressure of the unstable regime
can be reached even before attaining the
quantum string limit.

\vskip 1.5 cm

{\bf 4. Self-consistent string sources for the background equations.}

The field equations obtained from the low energy string effective
action for a spatially flat, homogeneous and isotropic metric and
dilaton background, with matter sources but vanishing dilaton
potential, can be written (in $d$ spatial dimensions) as [7,8,11]
$$
\dot {\fb} ^2- 2 \ddot {\fb} + d H^2 =0
$$
$$
\dot {\fb} ^2 -d H^2 = \rb e^{\fb}
$$
$$
2(\dot H- H\dot {\fb}) =\pb  e^{\fb} \eqno(4.1)
$$
Here $\fb$ is the duality-invariant shifted dilaton field,
$$
\fb =\phi - \ln \sqrt{|g|}= \phi -d \ln a \eqno(4.2)
$$
and the rescaled energy density and pressure
$$
\rb = \r \sqrt{|g|},~~~~~~\,\,\,\,\,\,\,\,\,\,
\pb = p\sqrt{|g|} \eqno(4.3)
$$
satisfy the conservation equation
$$
\dot {\rb} + d H \pb =0 \; . \eqno(4.4)
$$

As already discussed in [8,11], these equations can be integrated a
first time to give
$$
{1\over a}{d a\over dx}= {2 \Ga \over (x+x_0)^2- d \Ga ^2}\eqno(4.5)
$$
$$
{d \fb \over dx}= -{2(x+x_0)\over (x+x_0)^2- d \Ga ^2}\eqno(4.6)
$$
$$
\rb = {e^{\fb}\over 4\ell^2}[(x+x_0)^2- d \Ga ^2]\eqno(4.7)
$$
where $ \ell, x_0$ are integration constants, $x$ a new
(dimensionless) time parameter related to cosmic time by
$$
\rb={1\over \ell}{dx\over dt}\eqno(4.8)
$$
and
$$
\Ga (x)= \int ^x \ga (x') dx' \eqno(4.9)
$$
($\ga = p/\r = \pb /\rb$ as before). For any given time-dependent
equation of state $\ga (x)$ one can thus obtain, from
eqs.(4.5)-(4.7),
the general exact solution for $a(x), \phi(x)$ and $ \r (x)$, which
can
be eventually expressed in cosmic time through eq.(4.8).

If we are looking, however, for a simultaneous solution of the
background equations and of the string equations of motion, the
functional form of $\ga$ to be inserted into eq.(4.9) must be
consistent with the time-evolution of the given initial string
distribution. In particular, in the context of the fluid model
discussed
in the previous Sections, $\ga$ is to be obtained by solving the
evolution equation (2.6) or (2.11), and it will be expressed in
general
as a function of the background scale factor, $\ga= \ga (a)$. It
turns
out, therefore, that a more convenient variable to look for solutions
of the background equations (4.5)-(4.7), with a string gas as a
consistent matter sources, is $y=\ln a$ instead of $x$.

By defining a new function $\om (y)$ such that
$$
(x+x_0) = - 2\Ga {\om '\over \om}\eqno(4.10)
$$
(a prime denotes differentiation with respect to $y$), eq.(4.5)
implies then
$$
\om '' +{\Ga '\over \Ga} \om ' -{d\over 4} \om =0 \eqno(4.11)
$$
For any given $\Ga (y)$ this equation uniquely determines $\om (y)$,
which inserted into eq.(4.10) provides $x=x(y)$. By inverting such
relation we can eventually obtain $a= e^y= a(x)$. Moreover, the
dilaton equation (4.6) can be rewritten
$$
\fb '= -{2(x+x_0)\over (x+x_0)^2- d \Ga ^2}\left(dx\over dy\right)=
-  {x+x_0\over \Ga}= 2{\om  '\over \om} \eqno (4.12)
$$
so that $\fb$ is also completely determined in terms of $y$ as
$$
\fb (y) = \phi_0 +2 \ln \om (y) \eqno(4.13)
$$
($\phi_0$ is an integration constant). The corresponding energy
density is then, from eqs.(4.7), (4.10),
$$
\rb (y) = \left(\Ga \over \ell\right)^2 e^{\phi_0}
\left (\om^{' 2} -{d\over 4}\om^2\right)\eqno(4.14)
$$
We note, finally, that for any given $\Ga(y)$ the corresponding
time-evolution of the ratio $\ga= p/\r$ is determined by eqs.(4.9)
and (4.5) as
$$
\ga(y)= {d\Ga(y)\over dx} = {\Ga'\over 2\Ga}
\left ({\om^{' 2}\over \om ^2} -{d\over 4}\right)^{-1}\eqno(4.15)
$$

A simple, but important application of this method to obtain
self-consistent string-driven backgrounds, is  the linear
case $\Ga (y) \sim y$. This case corresponds, according to eqs.(4.8),
(4.9), to a phase in which $dx/dt= const$, and in which $\ga \sim
dy/dt = d(\ln a)/dt =H $, as suggested by the results of the previous
Section (see eqs.(3.20) and (3.4a) with $\b=3$). In this case the
solution of eq.(4.11) can be expressed in terms of modified Bessel
functions, and the full system of equations can be easily integrated.

On the other hand, a background with $dx/dt\sim \rb = const$
corresponds to an exact solution of the cosmological equations (4.1)
obtained by imposing, as initial conditions at $t=-\infty$, the
string
perturbative vacuum (flat spacetime, $\phi =-\infty$), with a small
but finite initial density of dust-like sources, $\r>0$, $p=0$ [11].
Such
a solution can be written explicitly, in the isotropic case, as
$$
a(t)=a_{0}|{t-2T\over t}|^{\pm 1/\sqrt
d},\,\,\,\,\,\,\,\,\,\,\,\,\,\,\,
e^{\fb} ={16 \ell^2e^{-\phi_0}\over |t(t-2T)|},
$$
$$
\rb ={1\over \ell}{dx\over dt}= {e^{\phi_0} \over 4\ell^2}= const,
\,\,\,\,\,\,\,\,\,\,\,\,\,\,p=0,
\,\,\,\,\,\,\,\, t< 0 \eqno(4.16)
$$
where $a_0,\phi_0$ and $T$ are positive integration constants (the
plus and minus sign in the exponent corresponds, for $t\ra 0_-$, to a
metric describing accelerated expansion and contraction,
respectively).

This background is certainly consistent with with a
string-driven evolution in the limit $t\ra -\infty$, and in
particular
for $|t|>>T$. Indeed, in this regime the metric becomes flat,
$$
a = const, \,\,\,\,\,\,\,\,\,\, \phi \sim -2\ln (-t),
,\,\,\,\,\,\,\,\,\,
\r = const \eqno(4.17)
$$
so that there are no horizons, all strings are stable and are thus
represented by a pressureless stress tensor. When
 $|t|\sim T$, however, the curvature scale begins to increase, and
all strings progressively enter the non-oscillating unstable regime.
The  time variation of the ratio $p/\r$ has then
to be taken into account, in
general, for an exact string-driven solution which may describe the
background evolution away from the string perturbative vacuum,
consistently with the string equations of motion. What is remarkable,
however, is that the simple background (4.16) may remain a good
zeroth-order approximation to the exact string-driven solution, if
the
duration of the accelerated regime (from $T$ to $\la_s$) is long
enough to satisfy the phenomenological constraints on inflation [11].

By adopting an iterative
procedure, let us assume indeed the solution (4.16) to be a
zeroth-order approximation, and let us compute the first-order
corrections by inserting that solution into the string
evolution equation (2.6), in order to obtain the corresponding
value of $\ga (t)$. We shall take for simplicity an initial string
distribution with $\b=3$, but our final conclusion turns out to be
independent of the particular choice of $\b$. Following the
procedure illustrated in Section 3 we find, for $|t|>>T$, that $\xi
\sim
t^2$, so that $\pi \a' E_u= \Lambda^2 D^{-1}$. As a consequence we have,
from eq.(2.10),
$$
\ga(t)=-{1\over d} H \la_s
\eqno(4.18)
$$
Note that $\ga$ goes  to zero like $-H$ as $t\ra -\infty$, in
agreement with our previous results.

We now insert this expression of $\ga(t)$ into the right-hand side of
the background equations (4.5)-(4.7), by recalling that for the
pressureless background (4.16) one has, to zeroth-order,
$$
\Ga^{(0)}= \pm {T\over 4\ell \sqrt d}e^{\phi_0}
\eqno(4.19)
$$
Therefore, to next order,
$$
\Ga=\Ga^{(0)}+\int _{-\infty}^x \ga(x^\pr)dx^\pr =\Ga^{(0)}-
{\la_s e^{\phi_0}\over d 4\ell}
\int_{-\infty}^t  H(t')dt'=
$$
$$
=\Ga^{(0)}\left( 1-{\la_s\over dT}\ln \left|t-2T \over
t\right|\right)
\eqno(4.20)
$$
According to our iterative approach, the integration
of eqs. (4.5)-(4.7) with the new expression (4.20) for $\ga$
provides a first-order approximation to the background
fields $a(t)$, $\phi(t)$.  The first-order corrections to (4.16)
due to a non-vanishing string effective pressure
are certainly negligible in the regime $|t|>>T$.
However, as clearly shown by eq. (4.20), the corrections may remain
small over the whole time range
if $T>>\la_s$.

In connection with this last point, it is important to note that the
ratio $T/\la_s$ measures the duration of the inflationary phase
associated to the solution (4.16). For $|t|<T$ the metric background
describes indeed a dilaton-dominated, accelerated evolution, which
can be approximated by
$$
a(t)\sim (-t)^{\mp 1/\sqrt d} \eqno(4.21)
$$
During such a phase the event horizon (1.5) shrinks linearly, and the
ratio $r$ of the proper size of a causally connected region to the
proper size of the horizon grows in time like $r(t)\sim (-t)^{-1\mp
1/\sqrt d}$, for $t\ra 0_-$. On the other hand, the horizon problem
of
the standard cosmological model is solved if the growth of $r(t)$,
for
$|t|$ ranging from $T$ to $\la_s$, is large enough to compensate the
decrease of $r(t)$ in the subsequent phase of expanding horizons,
down to the present time $t_0$. This requires [11], by assuming that
the end of inflation at $t=-\la_s$ is followed by the standard
radiation-dominated and matter-dominated evolution,
$$
\left(\la_s\over T\right)^{-1\mp 1/\sqrt d} \gaq
10^{30} \left({\ell}_p \over \la_s \right)^{1/2} \eqno(4.22)
$$

We can thus conclude that if the integration constant $T$ is chosen
in
such a way that the solution (4.16) may describe a
phenomenologically interesting inflationary background, then
$T>>\la_s$. In particular, for $\la_s \sim {\ell}_p$,
$$
 T \gaq
10^{30\sqrt d / (\sqrt d \pm 1)} \la_s \eqno(4.23)
$$
In that case, according to eq.(4.20), the background (4.16) also
represents a good zeroth-order approximation to an exact
simultaneous solution of the background equations and of the string
equations of motion, having the string perturbative vacuum as initial
condition.
\vfill\eject
{\bf 5. Conclusion.}

In this paper we have studied the number distribution of a classical
string network, evolving in a background in which the causal
correlation length $H^{-1}$ shrinks to a minimum and then
re-expands. Such kind of background emerges naturally from the
low energy string effective action when looking for cosmological
models based on the string perturbative vacuum as initial condition.

In such a background, the string proper size $L$ is constant outside
the horizon, while it varies like the scale factor outside the
horizon.
As a consequence, the number distribution $n(L)$ varies in time,
leading to a variation of the effective equation of state associated
to the string network. We have presented two possible approaches
to compute the ratio $\ga=p/\r$ in terms of the initial distribution
and of the background scale factor. In both cases we have checked
that, for a test string distribution, $\ga$ goes
to zero with a power-like behaviour  as $|H|^{-1}$ increases;
moreover, $\ga$ ranges over negative values when the
metric describes accelerated expansion, and over positive values
when the metric describes accelerated contraction. In the first case,
the transition from shrinking to expanding horizons is associated to
a
transition from decreasing to increasing pressure. The pressure,
however, stays always negative, unless quantum effects (such as
string decay into radiation) are included, which would eventually
bring the equation of state to that of relativistic matter, $\ga =
1/d$.

We have discussed, finally, a possible method to look for
simultaneous solutions of the background equations and of the
string propagation equation. We have shown that the simple
dust-like solution can approximate, self-consistently, the exact
string driven background describing the evolution from the initial
string perturbative vacuum to a final highly curved, strong coupling
regime. The corrections to such a zeroth-order approximation keep
small over the whole temporal range, provided the time extension
$T$ of the inflationary regime is much larger than one when
measured in string units, $T/\la_s >> 1$.
\vfill\eject

\centerline{\bf References.}

\item{1.}H. J. De Vega and N. S\'anchez, Phys. Lett. B197 (1987) 320.

\item{2.} N. S\'anchez and G. Veneziano, Nucl. Phys. B333 (1990) 253.

\item{3.}M. Gasperini, N. S\'anchez and G. Veneziano, Int. J. Mod.
Phys.
A6 (1991) 3853; Nucl. Phys. B364 (1991) 365;

Nguyen Suan Han and G. Veneziano, Mod. Phys. Lett A6 (1991) 1993.

\item{4.}M. Gasperini, Phys. Lett. B258 (1991) 70.

\item{5.}N. Turok and P. Bhattacharjiee, Phys. Rev. D29 (1984) 1557;

N. Turok, Phys. Rev. Lett. 60 (1988) 543.

\item{6.}G. Veneziano, Europhys. Lett. 2 (1986) 133.

\item{7.} G. Veneziano, Phys. Lett. B265 (1991) 287.

\item{8.}M. Gasperini and G. Veneziano, Astropart. Phys. 1 (1993)
317;
Mod. Phys. Lett. A8 (1993) 3701.

\item{9.}H. J. De Vega and N. S\'anchez, Phys. Rev. D50 (1994) 7202.

\item{10.}R. Brustein and G. Veneziano, Phys. Lett. B329 (1994) 429.

\item{11.} M. Gasperini and G. Veneziano, Phys. Rev. D50 (1994) 2519.

\end